\documentstyle[preprint,aps]{revtex}
\draft
\begin{document}
\title {\bf NEUTRINO EMISSIVITY OF DENSE STARS }
\author{\bf Sanjay K. Ghosh, S. C. Phatak and P. K. Sahu}
\address{Institute of Physics, Bhubaneswar-751005, INDIA.}
\footnotetext {$\mbox{}^*$ E-mail:
sanjay/phatak/pradip@iopb.ernet.in}
%\date{today}
\maketitle
\begin{abstract}
The neutrino emissivity of compact stars is investigated
in this work. We consider stars consisting of nuclear as well as
quark matter for this purpose. Different models are
used to calculate the composition of nuclear and quark matter
and the neutrino emissivity. Depending on the model
under consideration, the neutrino emissivity of nuclear as well as quark
matter varies over a wide range. We find that for nuclear
matter, the direct URCA processes are allowed for most of the
relativistic models without and with strange baryons, whereas
for the nonrelativistic models this shows a strong dependence on the type
of nuclear interaction employed. When the direct URCA processes are
allowed, the neutrino emissivity of hadronic matter is larger than that
of the quark matter by several orders of magnitude.
We also find that the neutrino emissivity departs from $T^6$ behavior
when the temperature is larger than the difference in the Fermi
momenta of the particles, participating in the neutrino-producing
reactions.
\vskip 0.4in
\end{abstract}
\vfill
\eject
\newpage
\section{{\bf Introduction}}
The neutrino emissivity of compact stars, such as neutron stars,
has been a subject of interest for quite some time. When
collapsed stellar core is formed after the supernova explosion,
it initially cools by emitting radiation till the temperature
falls to about $10^9$K or $\sim$0.1 MeV. Below this temperature,
cooling by radiation emission is not efficient and the star is
expected to cool by emitting neutrinos. The neutrino emission
occurs by the so-called URCA processes ( $n \rightarrow p + e^-
+ \overline \nu_e $ and $ p + e^- \rightarrow n + \nu_e $ ) and
the energy-momentum conservation requires that the proton Fermi
momentum ( $p_p^F$ ) should be larger than $0.5 \times p_n^F$.
This implies that the proton fraction in the neutron star must
be larger than $1/9$\cite{Pethick92}.
Initially, it was thought that such high proton fraction will
not be present in neutron stars and one has to find some other
mechanism of neutrino emission. One of the ways of getting around this
problem is to consider the so-called modified URCA
processes\cite{Friman79}. In these processes, the neutrino
emission occurs in the presence of another particle ( a nucleon )
which helps in satisfying the energy-momentum conservation.
However, the neutrino emissivity due to the modified URCA processes is
extremely small \cite{Glen80}-\cite{Riper81} to explain the
cooling rates of
these stars.  Later, it was observed that, if the star contains
pion\cite{Maxwell77} or kaon condensate\cite{Brown87} or if the
star consists of
quark matter\cite{Iwamoto80}-\cite{Datta88}, the direct URCA
processes are allowed and
the neutrino emissivity is substantially larger ( $\sim$ order
of magnitude ) than that due to modified URCA processes.
Recently, some authors have found that, for certain models
of the nuclear matter\cite{Lattimer91}, the proton fraction could be
larger than the critical value above which the direct URCA
processes are allowed. In such a situation, the neutrino emissivity of
the star could be large and one can explain the
cooling rates in terms of the standard neutron star models. This
explanation, however, depends on the assumption of the nuclear
interactions. In particular, it depends on the
nature of the three-body force and its isospin
dependence\cite{Wiringa88}.
\par
The purpose of the present work is to calculate the neutrino emissivity
of nuclear
and quark matter using different models. For nuclear
matter, we use nonlinear Walecka model,
derivative scalar coupling model, chiral
$\sigma$ model and
nonrelativistic models, whereas, for the quark matter, we use the
MIT bag model and the chiral color dielectric (CCD) model. One of the
reasons behind this calculation is to determine the nuclear
equations of state in which the direct URCA processes
are allowed. This is of some importance, since
the neutrino emissivity due to direct URCA processes dominates
when these are allowed. We also want to explore the dependence
of the neutrino emissivity on different quark models. In our
earlier calculations of neutrino emissivity of the quark
matter\cite{GPS1}, we found that, for certain values
of Fermi momenta, the approximate neutrino emissivity formula of
Iwamoto\cite{Iwamoto80} fails.
In that work\cite{GPS1}, we were able to determine an empirical
formula which is able to reproduce the calculated neutrino emissivity
over a wide range of quark matter densities and temperatures.
In the present calculation, we want to investigate, if a similar situation
exists for nuclear matter also.
\par
The results of our calculation can be summarised as follows. We
find that for all the relativistic models considered here, the direct URCA
processes are forbidden if the nuclear density is below a
certain value. In other words, for all the models, the
proton fraction is not high enough at lower nuclear densities. The
neutrino emissivity,  for all the models without
 strange baryons, falls rapidly at higher densities (more than 5
times nuclear matter densities ).
But when the strange baryons
are included, the neutrino emissivity due to direct URCA process at
higher densities varies very slowly with baryon densities. Thus, the
appearance of strange baryons depletes the neutron fraction and
the direct URCA processes are allowed, even though the proton
fraction is not large. We find that the
reactions involving strange baryons give a significant
contribution to the neutrino emissivity. As for
the quark matter, we find a large dependence of the neutrino emissivity on
the models used, to calculate quark matter equation of state.
However, when the
direct URCA processes in the nuclear matter are allowed, the
neutrino emissivity of the nuclear
matter is larger than that of the quark matter at corresponding
nuclear density. We also find that the calculated neutrino
emissivity departs from $T^6$\cite{Iwamoto80,Lattimer91}
behaviour for a certain range of Fermi momenta of the
constituents. This happens for nuclear as well as for quark
matter. Furthermore, we are able to fit the calculated
neutrino emissivity with a simple universal formula. This clearly implies
that the departure from $T^6$ behavior of the neutrino emissivity is
kinematical in origin.
\par
The paper is organised as follows. In Section II, we briefly describe
the models used in the calculation of nuclear and quark equations
of state. In Section III, the neutrino emissivity formulae are presented.
Finally, the results are discussed in Section IV.
\section{{\bf The Models}}
\subsection{Nuclear Models}
Four different models, the nonlinear Walecka model\cite{Kapusta90,GPS2},
derivative scalar coupling model\cite{Zimanyi90,Glendenning92},
chiral $\sigma$ model\cite{Sahu93} and nonrelativistic
model\cite{Wiringa88} have been used to calculate the equation
of state of the neutron matter. Of these, the first two have
been extensively used in nuclear structure
calculations\cite{Serot86,Zimanyi90} and have been able to
reproduce the properties of nuclei over a wide range of the
periodic table. This probably ensures that one has correct
nuclear equation of state near the nuclear matter density. The
chiral $\sigma$ model has been used as another
model for the nuclear equation of state. This model has been
used to calculate neutron star properties\cite{Sahu93}.
One important fact about this model is that nonlinear terms can
give rise to the three body forces, which is important in the
equation of state at high densities.
For all these relativistic models,
the nuclear equation of state is calculated by adopting the mean
field ansatz. This is in contrast to the nonrelativistic models,
where actual interaction between the constituents are considered.

{\bf The Nonlinear Walecka Model: (NW)} The Lagrangian density of the
nonlinear Walecka model\cite{GPS2} is given by,
\begin{eqnarray}
{\cal L}(x)~=\sum_{i} \bar \psi_{i} (i\gamma^{\mu}\partial_{\mu}
- m_i +g_{\sigma i}
\sigma+ g_{\omega i} \omega_{\mu} \gamma^{\mu}-
g_{\rho i} \rho^{a}_{\mu} \gamma^{\mu} T_{a} ) \psi_{i}
-{1 \over {4}} \omega ^{\mu \nu} \omega_{\mu \nu} \nonumber \\
+{1 \over {2}} m^{2}_{\omega} \omega_{\mu} \omega^{\mu}
+ {1 \over {2}} ( \partial_{\mu} \sigma \partial^{\mu} \sigma-
m^{2}_{\sigma} \sigma^{2})
- {1 \over {4}} \rho^{a}_{\mu \nu} \rho^{\mu \nu}_{a} +
{1 \over {2}} m_{\rho}^{2} \rho^{a}_{\mu} \rho^{\mu}_{a} \nonumber \\
 - {1 \over {3}} bm_{N} (g_{\sigma N} \sigma)^{3} -
{1 \over {4}} c( g_{\sigma N} \sigma )^4 +
\sum_{l}\bar\psi_{l}(i\gamma^{\mu}\partial_{\mu}-m_l)\psi_{l}
\label{eq:walL}
\end{eqnarray}
\noindent The Lagrangian in eq.(\ref{eq:walL})above includes
nucleons, $\Lambda$ and
$\Sigma^-$ hyperons (denoted by subscript $i$), electrons and muons
(denoted by subscript $l$) and $\sigma$, $\omega$ and $\rho$ mesons
(given by $\sigma$, $\omega^\mu$ and $\rho^{a,\mu}$ respectively).
The Lagrangian includes cubic and quartic self-interactions of
the $\sigma$ field. The meson fields interact with baryons through
linear coupling and the coupling constants are different for nonstrange
and strange baryons.
\par
In the presence of baryons, the mesons develop nonzero vacuum expectation
values ($\bar \sigma$, $\bar \omega$ and $\bar \rho^a$ respectively).
Assuming that the baryon densities are uniform, one finds that
the time components of $\bar \omega$ and  $\bar
\rho^3$, in addition to $\bar \sigma$, are nonzero. One can then
define effective masses ($\bar m_i$)
and chemical potentials ($\bar \mu_i$) for the baryons as,
\begin{eqnarray}
\bar m_i= m_i - g_{\sigma i} \bar \sigma
\end{eqnarray}
\noindent and
\begin{eqnarray}
\bar \mu_i = \mu_i - g_{\omega i} \bar \omega - I_3 g_{\rho N}
\bar \rho^{3}, \label{eq:chemical}
\end{eqnarray}
\noindent where $I_3$ is the value of the z-component of the
isospin of baryon $i$.
The Fermi momenta ($k_i$) and number densities ($n_i$) of the baryons
are given by $k_i~=~{\sqrt{\bar \mu_i^2 - \bar m_i^2}}$ and
$n_i~=~{\frac {k_i^3}{3 \pi^2}}$.
For leptons, the Fermi momenta and number densities are given by
$k_l~=~\sqrt{\mu_l^2-m_l^2}$ and $n_l~=~\frac {k_l^3}{3 \pi^2}$.
\par
The parameters of the NW model are meson-baryon coupling constants,
meson masses and the coefficients of the cubic and quartic
self-interactions of $\sigma$ meson ( $b$ and $c$ respectively).
The $\omega$ and $\rho$ meson masses have been chosen to
be their physical masses. Of the rest of the parameters, the
nucleon-meson coupling constants ( ${g_{\sigma} \over {m_{\sigma}}}$,
${g_{\rho} \over {m_{\rho}}}$ and ${g_{\omega} \over
{m_{\omega}}}$ respectively ) and the coefficients of cubic and
quartic terms of the $\sigma$ meson self interaction ( b and c
respectively) are determined by fitting the nuclear matter
properties ( the binding energy/nucleon ($-16
MeV$ ) and baryon density ($0.15fm^{-3}$ ), symmetry energy
coefficient ($32.5 MeV$ ), Landau mass ($0.83 m_N$) and nuclear
incompressibility ($250- 300  MeV$) ). The coupling constants of the
hyperon-meson interactions and are not well
known. These cannot be determined from nuclear matter properties,
since the nuclear matter does not contain hyperons. Furthermore,
properties of hypernuclei do not fix these parameters in
a unique way. In the literature, a number of choices have been
made. These are
(a) same as the nucleon-meson coupling
constants (Universal coupling) \cite{Glendenning85,Glendenning92}
,(b) $\sqrt{2/3}$ times the nucleon-meson coupling
constants\cite{Glendenning85,Glendenning92},
(c) $1/3$ times the nucleon-meson coupling
constants \cite{Kapusta90}
and (d) $2/3$, $2/3$ and $1$ times the nucleon- meson coupling for $\sigma$,
$\omega$ and $\rho$ mesons respectively\cite{GPS2}.
\par
We have used the above mentioned choices for meson- strange
baryon couplings to investigate the neutrino emissivity.

{\bf Derivative Scalar Coupling Model (DSC model):} In this case, the
Lagrangian is given by\cite{Zimanyi90}
\begin{eqnarray}
{\cal L}(x)~=\sum_{i} \bar \psi_{i} (i\gamma^{\mu}\partial_{\mu}
- m_i + g_{\omega i} \omega_{\mu} \gamma^{\mu}-
g_{\rho i} \rho^{a}_{\mu} \gamma^{\mu} T_{a} ) \psi_{i}
-{1 \over {4}} \omega ^{\mu \nu} \omega_{\mu \nu} \nonumber \\
+{1 \over {2}} m^{2}_{\omega} \omega_{\mu} \omega^{\mu}
+ {1 \over {2}} ( \partial_{\mu} \sigma \partial^{\mu} \sigma-
m^{2}_{\sigma} \sigma^{2})
- {1 \over {4}} \rho^{a}_{\mu \nu} \rho^{\mu \nu}_{a} +
{1 \over {2}} m_{\rho}^{2} \rho^{a}_{\mu} \rho^{\mu}_{a} \nonumber
\\
+\sum_{i}g_{\sigma}\bar\psi_{i} \sigma \psi_{i} (1+g_{\sigma}\sigma/M)
+ \sum_{l}\bar\psi_{l}(i\gamma^{\mu}\partial_{\mu}-m_l)\psi_{l}
\label{eq:dscL}
\end{eqnarray}
\noindent where  $\omega_{\mu}$ represents the  vector meson field,
$m_{\sigma}$ and $m_{\omega}$ are the masses of the scalar and vector fields,
and $F_{\mu\nu}=\partial_{\mu}\omega_{\nu}-
\partial_{\nu}\omega_{\mu}$. The summation indices $i$ and $l$
stand for fermions and leptons respectively. The DSC Lagrangian
differs from the original $\sigma$ - $\omega$ model\cite{Walecka74} in the
baryon- $\sigma$ meson coupling term
($= g_{\sigma}\bar\psi \psi \sigma$ for the $\sigma$ - $\omega$
model), but it has also two parameters: $g_{\sigma}$ and
$g_{\omega}$ like the original model\cite{Walecka74}.
\par
 From the Lagrangian (\ref{eq:dscL}), the following definition of (density
dependent) effective nucleon mass ($m^{*}$) is suggested :

\begin{equation}
m^{*} =  M/(1+g_{\sigma}\sigma/M)
\end{equation}

{\bf Chiral Sigma Model (CS model): }
This model includes $\sigma$, $\omega$ and $\pi$ fields. In
addition, $\rho$ meson is included in the Lagrangian to
reproduce the symmetry energy of the nuclear matter correctly.
The Lagrangian for an SU(2) $\times$ SU(2) chiral sigma model that
includes (dynamically) an isoscalar vector field ($\omega_{\mu}$) is
\begin{eqnarray}
{\cal L}~(x) =  \frac{1}{2}\big(\partial_{\mu} \overrightarrow{\pi} .
\partial ^{\mu} \overrightarrow{\pi} + \partial_{\mu} \sigma
\partial^{\mu} \sigma\big) - \frac{\lambda}{4}\big(\overrightarrow{\pi} .
\overrightarrow{\pi} +\sigma^{2} - x^2_o\big)^2\nonumber\\
 - \frac{1}{4} F_{\mu\nu} F_{\mu\nu} + \frac{1}{2}g_{\omega}
\big(\sigma^2 + \overrightarrow{\pi}^2\big) \omega_{\mu}
\omega^{\mu}
 + g_{\sigma} \bar{\psi} \big(\sigma + i\gamma_5 \overrightarrow{\tau}
. \overrightarrow{\pi}\big) \psi \nonumber \\ + \bar\psi \big(
i\gamma_{\mu}\partial^{\mu} - g_{\omega}\gamma_{\mu}
\omega^{\mu}\big) \psi
-\frac {1}{4}G_{\mu\nu}G^{\mu\nu}
+\frac{1}{2}m^2_{\rho}\overrightarrow{\rho_{\mu}}.
\overrightarrow{\rho}^{\mu} \nonumber \\
-\frac{1}{2}g_{\rho}\bar\psi(\overrightarrow{\rho}_{\mu}.
\overrightarrow{\tau}\gamma^{\mu}) \psi
+ \sum_{l}\bar\psi_{l}(i\gamma^{\mu}\partial_{\mu}-m_l)\psi_{l}
\label{eq:chiL}
\end{eqnarray}

where $F_{\mu\nu} \equiv \partial_{\mu} \omega_{\nu} - \partial_{\nu}
\omega_{\mu}, \psi$ is the nucleon isospin doublet,
$\overrightarrow{\pi}$ is the pseudoscalar pion field and $\sigma$ is the
scalar field.  The vector field $\omega_{\mu}$ couples to the conserved
baryonic current $j_{\mu} = \bar{\psi} \gamma_{\mu} \psi$. The expectation
value $<j_o>$ is identifiable as the nucleon number density.
\par
The interactions of the scalar and the pseudoscalar mesons with the vector
boson generates a mass for the latter spontaneously by the Higgs
mechanism. The masses for the nucleon, the scalar meson and the vector
meson are respectively given by
\begin{eqnarray}
M = g_{\sigma} x_o \nonumber\\
m_{\sigma} = \sqrt{2\lambda} x_o \nonumber\\
m_{\omega} = g_{\omega} x_o.
\end{eqnarray}
where $x_{0}$ is the vacuum expectation value of sigma field.
\par
With $x = (<\sigma^2 + \overrightarrow{\pi}^2>)^{1/2}$, the
the effective mass of the nucleon is $M^{\star} \equiv yM$,
where $y=x/x_{0}$. For both DSC and CS model, the chemical
potential of the baryon can be related to the fermi momentum
through a similar relation as eq.(\ref{eq:chemical}). The
nucleon- meson coupling constant are determined by fitting the
nuclear matter properties, binding energy (-16.3 MeV),
saturation density (0.153 $fm^{-3}$) and symmetry energy
coefficient (32.5 MeV). The incompressibility in the above two
models are 225 MeV and 700 MeV respectively.
\par
{\bf Nonrelativistic models: } Here we have used the models as
proposed by Wiringa et al.\cite{Wiringa88} by combining different
two- nucleon and three- nucleon potentials. In particular three
different choices have been considered.
\begin{itemize}
\item Argonne $v_{14}$ (AV14) and Urbana VII (UVII) three
nucleon potential,
\item Urbana $v_{14}$ (UV14) two nucleon potential and UVII,
\item UV14 and three nucleon interaction (TNI) model of Lagaris
and Pandharipande\cite{Lagaris}.
\end{itemize}
\par
AV14 and UV14 have identical structure and can be written as a
sum of 14 operator components. The main difference between these
two models come from the strength of short range tensor force.
The UV14 does not have short range tensor components which results in a
weak tensor force that vanishes at the origin. The AV14 tensor
force is finite at the origin and at intermediate distance looks
like Paris potential\cite{Lacombe}. The three nucleon potential
potential UVII combines a long range two pion exchange part and an
intermediate range repulsive part. Out of the three
different combinations, $AV14~+ ~UVII$, $UV14~+ ~UVII$ and
$UV14~+ TNI$ considered, only the second combination produce
enough proton fraction to have non- zero neutrino emissivity.
\par
In order to have some idea on the expected behaviour of
different models, let us study the proton fractions for
different parameters as well as different models. In fig. 1,
 we have plotted the
proton fraction for nonlinear Walecka model with
different hyperon couplings and incompressibility. It is evident from this
figure that, for all the parameter sets, the required proton
fraction is attained over a certain density ($\sim 2\rho_{0}$,
$\rho_{0}$ is the nuclear matter density ).
A change in incompressibility from 300 MeV
to 350 MeV does not give much change in the proton fraction. On
the other hand, the variation of hyperon couplings give a large
change in the proton fraction. This indicates that the
effect of hyperon couplings on neutrino emissivity will be much larger
than that of the incompressibilities.
\par
Fig.2. gives a comparison of proton fraction between different
models. It shows a strong density dependence of the proton
fraction on models. Below a certain density ($\sim 0.2
fm^{-3}$), none of
these models satisfy the limit and neutrino emissivity will be
zero. We also find that nonlinear Walecka model without hyperons
yield larger proton fraction at higher density ($> 0.8 fm^{-3}$). Also,
with hyperons required proton fraction is attained at earlier
baryon densities than without hyperons.
However, the nonrelativistic model yield lowest proton fraction
throughout the density range.
\par
The proton fraction also depends on the symmetry energy
coefficients. We have varied the symmetry energy coefficient
from 28- 38 MeV for DSC and CS models. With increase in symmetry
energy, the required proton fraction for neutrino emissivity in
direct URCA process, is attained at lower baryon density.
\subsection{Quark Models}
We have used CCD model and bag model\cite{GPS1} to study the
neutrino emissivity.
The colour dielectric model is based on the idea of Nielson
and Patkos \cite{Nielson82}. In this model, one generates the
confinement of quarks and gluons dynamically through the
interaction of these
fields with scalar field. Here, we have used chiral extension of
this model to study the quark matter neutrino emissivity. The CCD model has
already been used to study static properties of baryons \cite{SSahu92},
properties of quark matter at finite density and temperature
\cite{Ghosh92,Ghosh93}, hadron-quark phase transition and properties of dense
stars \cite{GPS2,Ghosh93a}. In CCD quark matter, we assume that meson ($\phi$)
expectation value is zero i.e. $<\phi>= 0$ where as
$<\phi^{2}>\neq 0$. So, the Lagrangian is rewritten in terms of
$<\phi^{2}>$. The quark mass becomes density dependent. In this
model, with increase in density quark mass decreases and drops to
about $1/4$th of its initial value around density $2 fm^{-3}$.
This density is defined as the critical density for chiral
transition in our model. There are five parameters in CCD model; bag
parameter $B$, scalar field potential parameter $\alpha$, $u$
and $d$ quark mass, strange quark mass and strong coupling
constant $\alpha_{s}$. These input parameters are obtained by
fitting the baryonic masses. In the present paper, we have discussed
the results for the parameter set: $B^{1/4}= 152.1MeV$, $m_{q(u,d)}=
91.6MeV$, $m_{q(s)}= 294.9MeV$, $\alpha=36$ and strong coupling
constant $\alpha_{s}= 0.08$. In the bag model, the neutrino emissivity is
calculated with $m_{u}= m_{d}=0$, $m_{s}= 150 MeV$ and
$\alpha_{s}= 0.08$. For both CCD and bag model, we consider
interaction upto first oreder in $\alpha_{s}$.
\section{{\bf Emissivity Formulae}}
In general, the Lagrangian density for URCA processes in
the current current interaction form is written as\cite{Commins83},
\begin{eqnarray}
{\cal L}(x)={\frac{G_{F}}{\sqrt{2}}}l_{\mu}(x){\cal
I}^{\mu}(x)~+~H.C
\label{eq:lsalam}
\end{eqnarray}
where the weak coupling constant $G_{F}= 1.435\times 10^{-49}erg~
cm^3$ and $l_{\mu}$ and ${\cal I}^{\mu}$ are the leptonic and
hadronic weak currents respectively.
\begin{eqnarray}
l_{\mu}(x)=\bar {e}\gamma_{\mu}(1-\gamma_{5})\nu_{e}+\bar
{\mu}\gamma_{\mu} (1-\gamma_{5})\nu_{\mu}+............+h.c.
\label{eq:weak}
\end{eqnarray}
\begin{eqnarray}
{\cal I}_{\mu}(x)={\bar {\psi}}_{1}\gamma_{\mu}(A-B \gamma_{5})
\psi_{2}~+~h.c.
\label{eq:strong}
\end{eqnarray}
where $h.c.$ stands for hermitian conjugate. For quarks $A= B=1$
and for baryons the value of $A$ and $B$ depends on the specific nature
of the particle. Using the above Lagrangian one can calculate the
neutrino emissivity $\epsilon$,
\begin{eqnarray}
\epsilon= g\int \{\Pi_{i=1}\frac{d^{3}p_{i}}{(2\pi)^3}\}
E_{\nu}W_{fi}F(p_1,p_2,p_e)
\label{eq:genems}
\end{eqnarray}
where $i=1,2,e,\nu$, $E_{\nu}$ is the energy of the neutrino and
$g$ is the degeneracy factor.
The transition rate is
\begin{eqnarray}
W_{fi}= \frac{(2\pi)^4 \delta^{4}(P_{in}-P_{fn})|M|^{2}}{\Pi_{i}2E_{i}}
\label{eq:trans}
\end{eqnarray}
where $P_{in}$ is the sum of the initial momenta, $P_{fn}$ is the sum of
the final momenta, $E_i$ is the energy of the $i$th particle
and $|M|^{2}$ is the squared invariant amplitude averaged over
initial spins and summed over final spins. The symbol $F(p_{1},p_{2},p_{e})$
for the reactions $1\rightarrow 2+e^{-}+\bar{\nu_{e}}$ and
$2+e^{-}\rightarrow 1+\nu_{e}$ are
\begin{eqnarray}
F_{d}(p_{1},p_{2},p_{e})=n(p_1)(1-n(p_2))(1-n(p_e))
\label{eq:fpp1}
\end{eqnarray}
and
\begin{eqnarray}
F_{r}(p_{1},p_{2},p_{e})=(1-n(p_1))n(p_2)n(p_e)
\label{eq:fpp2}
\end{eqnarray}
respectively, where
\begin{eqnarray}
n(p)= \frac{1} {1~+~e^{(E-\mu)/T}} .
\end{eqnarray}
\subsection{Hadronic matter}
In hadronic matter, several weak decays may contribute to the
neutrino emission as given below\cite{Prakash92}.
\begin{eqnarray}
n\rightarrow p~+~e^{-}~+~\bar\nu_{e}; ~~ p~+~e^{-}\rightarrow
n~+~\nu_{e} \nonumber \\
A=cos\theta_{c}; ~~ B=(F~+~D)cos\theta_{c}  \\  \label{eq:npe}
\Lambda\rightarrow p~+~e^{-}~+~\bar\nu_{e}; ~~
p~+~e^{-}\rightarrow \Lambda~+~\nu_{e} \nonumber \\
A=-3sin\theta_{c}/\sqrt{6}; ~~ B=-\frac{(3F~+~D)}{\sqrt{6}}sin\theta_{c}  \\
\label{eq:lpe}
\Sigma\rightarrow n~+~e^{-}~+~\bar\nu_{e}; ~~
n~+~e^{-}\rightarrow \Sigma~+~\nu_{e} \nonumber  \\
A=-sin\theta_{c}; ~~ B=-(F~-~D)sin\theta_{c}  \\
\label{eq:sne}
\Sigma\rightarrow \Lambda~+~e^{-}~+~\bar\nu_{e}; ~~
\Lambda~+~e^{-}\rightarrow \Sigma~+~\nu_{e} \nonumber \\
A=0; ~~ B=\sqrt{\frac{2}{3}}D cos\theta_{c}
\label{eq:sle}
\end{eqnarray}
where $F=0.427$ and $D=0.823$. The angle $\theta_{c}$ is the
Cabbibo angle and $\cos\theta_{c}=0.948$. The $|M|^2$ for the
above processes is given by
\begin{eqnarray}
|M|^2={\frac{1}{2}} [64(A^2+B^2)\{(p_{2} . p_{e})(p_{1} . p_{\nu}) +
(p_{2} . p_{\nu})(p_{1} . p_{e}) \} \nonumber \\
+64AB\{2(p_{2} . p_{e})(p_{1} . p_{\nu}) -
2(p_{2} . p_{\nu})(p_{1} . p_{e}) \} \nonumber \\
+64(A^2-B^2)m_{1} m_{2}(p_{e} . p_{\nu}) ] \label{eq:hmat}
\end{eqnarray}
Using the above matrix element, one can calculate the neutrino emissivity
for both direct and reverse processes.
\begin{eqnarray}
\epsilon_{(1 \rightarrow 2+e^{-}+\bar{\nu_{e}})}&=&\left [ \int
p_2p_{e}{p_{\nu}}^{2}p_{1} dp_2dp_{e}dp_{\nu}dp_{1} \right. \nonumber \\
&&\left\{C_{1}\times \int_{max \{|p_2-p_{e}| , |p_{1}-p_{\nu}| \} }^{min \{
|p_2+p_{e}| , |p_{1}+p_{\nu}| \}}dP \; \;
(1 - { p_1^2 + p_{\nu}^2 - P^2 \over 2E_1 E_{\nu}}) \; \; (1
+ { p_{2}^2 + p_{e}^2 -P^2 \over 2E_{2}E_{e}}) \right. \nonumber \\
&&\left. +C_{2}\times \int_{max \{|p_1-p_{e}| , |p_{2}-p_{\nu}| \} }^{min \{
|p_1+p_{e}| , |p_{2}+p_{\nu}| \}}dP \; \;
(1 + { p_2^2 + p_{\nu}^2 - P^2 \over 2E_2 E_{\nu}}) \; \; (1
- { p_{1}^2 + p_{e}^2 -P^2 \over 2E_{1}E_{e}}) \right\} \nonumber \\
&&\left. -C_{3}m_{1}m_{2}\times {\int {{p_2}^{2}{p_{e}}^{3}{p_{\nu}}^{3}p_{1}
dp_2dp_{e}dp_{\nu}}\over {E_1E_2E_e}} \right]
\delta(E_2+E_{e}+E_{\nu}-E_{1}) F_{d}(p_1,p_2,p_e)
\label{eq:deps} \\
\epsilon_{(2+e^{-} \rightarrow 1+{\nu_{e}})}&=&\left[ \int
p_2p_{e}{p_{\nu}}^{2}p_{1} dp_2dp_{e}dp_{\nu}dp_{1}\right. \nonumber \\
&&\left\{C_{1}\times \int_{max \{|p_2-p_{e}| , |p_{1}-p_{\nu}| \} }^{min \{
|p_2+p_{e}| , |p_{1}+p_{\nu}| \}}dP \; \;
(1 + { p_1^2 + p_{\nu}^2 - P^2 \over 2E_1 E_{\nu}}) \; \; (1
+ { p_{2}^2 + p_{e}^2 -P^2 \over 2E_{2}E_{e}})\right. \nonumber \\
&&\left. +C_{2}\times \int_{max \{|p_1-p_{e}| , |p_{2}-p_{\nu}| \} }^{min \{
|p_1+p_{e}| , |p_{2}+p_{\nu}| \}}dP \; \;
(1 - { p_2^2 + p_{\nu}^2 - P^2 \over 2E_2 E_{\nu}}) \; \; (1
- { p_{1}^2 + p_{e}^2 -P^2 \over 2E_{1}E_{e}}) \right\} \nonumber \\
&&\left. -C_{3}m_{1}m_{2}\times {{\int {p_2}^{2}{p_{e}}^{3}{p_{\nu}}^{3}p_{1}
dp_2dp_{e}dp_{\nu}}\over {E_1E_2E_e}} \right]
\delta(E_2+E_{e}-E_{\nu}-E_{1}) F_{r}(p_1,p_2,p_e)
\label{eq:reps}
\end{eqnarray}
where $C_{1} =\frac{{G_{F}}^2(A+B)^2}{2(2\pi)^5} $,
$C_{2} =\frac{{G_{F}}^2(A-B)^2}{2(2\pi)^5} $  and
$C_{3} =\frac{2{G_{F}}^2(A^2-B^2)}{(2\pi)^5} $
\par
Similar decays involving $\mu^{-}$, instead of $e^{-}$ are also
possible. Here we have considered only $e^{-}$ channel, because
$e^{-}$ density will be much more than the $\mu^{-}$ and hence it will
contribute more.
In  URCA processes involving only one hyperon,
there is a change in strangeness, so the neutrino emission rate
which is proportional to $sin^{2}\theta_c$, is less than one
tenth of those for nucleonic URCA processes (${\propto
cos^{2}\theta_{c}}$). But in the processes involving only
hyperons, there is no change in strangeness and hence such reactions are
not Cabbibo suppressed.
\par
The nonrelativistic reduction of the matrix element
(eq.(\ref{eq:hmat})) can be obtained by neglecting the
baryon momenta and replacing baryon energy by corresponding
masses. The reduced matrix element becomes,
\begin{eqnarray}
|M|^2=\frac{1}{2}[64\{(A^2+3B^2)E_{e}E_{\nu} + (A^2-B^2)(p_{e} .
p_{\nu})\} m_{1} m_{2} ] \label{eq:hrmat}
\end{eqnarray}
The corresponding neutrino emissivity formula\cite{Prakash92} can be derived
from
eq.(\ref{eq:genems}) by performing the phase space integral using Fermi
liquid theory,
\begin{eqnarray}
\epsilon={\frac{457\pi}{10080}}
{G_{F}}^2C^2(A^2+3B^2)m_{1}m_{2}\mu_{e}T^{6} \label{eq:redeps}
\end{eqnarray}
\subsection{Quark matter}
In quark matter, the neutrino emission takes place due to
following processes.
\begin{eqnarray}
d\rightarrow u~+~e^{-}~+~\bar\nu_{e}; ~~
u~+~e^{-}\rightarrow d~+~\nu_{e}, \\
s\rightarrow u~+~e^{-}~+~\bar\nu_{e}; ~~
u~+~e^{-}\rightarrow d~+~\nu_{e} .
\end{eqnarray}
The matrix element for the quark URCA process is
\begin{eqnarray}
{|M|^2}_{d(s)}= 64cos^2\theta_{c}(sin^2\theta_{c})
[(p_{2} . p_{e})(p_{1} . p_{\nu})]
\label{eq:qmatr}
\end{eqnarray}
The neutrino emissivity for the quark URCA processes are given by
\begin{eqnarray}
\epsilon_{u \rightarrow d \; (u \rightarrow s)}(e^-) &=& C_{ud} (C_{us})\int
p_up_{e}{p_{\nu}^2}p_{d(s)} dp_udp_{e}dp_{\nu}dp_{d(s)}
\delta(E_u+E_{e}-E_{\nu}-E_{d(s)}) \nonumber \\
&& \times \;\; {1 \over e^{(E_u-\mu_u)/T}+1} \; \; {1 \over
e^{(E_e-\mu_e)/T}+1} \; \; {1 \over
e^{(\mu_{d(s)}-E_{d(s)})/T}+1} \nonumber \\
&& \times \int_{max \{|p_u-p_{e}| , |p_{d(s)}-p_{\nu}| \} }^{min \{
|p_u+p_{e}| , |p_{d(s)}+p_{\nu}| \}}dP \; \;
(1 + { p_u^2 + p_{\nu}^2 - P^2 \over 2 E_u E_{\nu}}) \; \; (1
+ { p_{d(s)}^2 + p_{e}^2 -P^2 \over 2E_{d(s)}E_{e}})
\label{eq:qreps} \\
\epsilon_{d \rightarrow u ( s \rightarrow u)}(e^-) &=& C_{du} (C_{su})\int
p_{d(s)}p_{u}p_{e}{p_{\nu}}^2 dp_{d(s)}dp_{u}dp_{e}dp_{\nu}
\delta(E_{d(s)}-E_{u}-E_{e}-E_{\nu}) \nonumber \\
&& \times \;\; {1 \over e^{(E_{d(s)}-\mu_{d(s)})/T}+1} \; \;
{1 \over e^{(\mu_{u}-E_{u})/T}+1} \; \; {1 \over
e^{(\mu_e-E_e)/T}+1} \nonumber \\
&& \times \int_{max \{
|p_{d(s)}-p_{\nu}| , |p_u-p_{e}| \} }^{min \{ |p_{d(s)}+p_{\nu}| ,
|p_u+p_{e}| \}} dP \; \;
(1 - { p_{d(s)}^2 + p_{\nu}^2 - P^2 \over 2E_{d(s)} E_{\nu}})
\; \; (1 - {P^2- p_u^2 + p_{e}^2 \over 2E_uE_{e}})
\label{eq:qdeps}
\end{eqnarray}
where
$~~C_{ud}
=C_{du}=\frac{6{G_{F}}^{2}cos^2\theta_{C}}{(2\pi)^5}$, $~~C_{su}=
C_{us}=\frac{6{G_{F}}^{2}sin^2\theta_{C}}{(2\pi)^5}$, $G_{F}$ is the weak
decay constant and $\theta_C$ is the Cabibbo
angle. Here $E_i$ and $\mu_i$ are energy and chemical potential of
the $i$-th species ($i=u,d,s,e$).
The single particle energy momentum relation, required to
evaluate neutrino emissivity for both hadronic and quark matter, is defined in
analogy with the relation between chemical potential and fermi
momentum.
\section{{\bf Results and Discussion}}
In the present neutrino emissivity calculation, we have
considered only the direct URCA processes for hadron and
quark matter. Also, the neutrino emissivity due to direct decay
process and its reverse, at chemical equilibrium, are taken to
be equal. The neutrino emission from charge neutral hadronic or
quark matter at chemical equilibrium, for small temperatures
($T\le 1MeV$ ), occur due to those fermions whose momenta lie
close to their Fermi surfaces. Therefore, the required
kinematical criterion for reaction to occur is that the momentum
conservation condition be satisfied for the fermion
momenta around the respective Fermi surfaces. For the
process $1\rightarrow 2~+~ e^{-}~ +~ \nu$, the momentum
conservation condition can be written as ${p_{2}}^{F}+
{p_{e}}^{F}- {p_{1}}^{F} =\Delta p > 0$.
\par
In fig. 3, we have plotted the density dependence of neutrino
emissivity at $T=0.5 MeV$ from different hadronic models which
are nonlinear Walecka model, derivative scalar coupling model
and chiral sigma model. The
quark matter neutrino emissivity using CCD and bag model are
also plotted in fig. 3. In CCD model, quarks
are massive ($m_{(u,d)}= 125MeV$, $m_s= 300MeV$). The $u$ and $d$
quarks masses decrease with increase in density due to non zero
$<\pi^2>$. On the other hand, $m_s$ remains unchanged as $<K^2>$ and
$<\eta^2>$ remain zero in the medium throughout the range of
densities considered \cite{Ghosh93}. We find that the
the neutrino emissivity in CCD model is higher ($\sim$ order of
magnitude) than the the bag model, due to the additive effect of
quark masses on the neutrino emissivity, the strong interaction coupling
constant being same ($\alpha_s$ is equal to $0.08$ for both CCD
and bag model where $\alpha_{s}={g_{s}}^2/{4\pi}$). Fig.3.
shows that upto density $1.0 fm^{-3}$, neutrino emissivity from
DSC and CS model matter is higher ($\sim$ factor of 2- 4) than
the other hadronic models. But beyond that only nonlinear Walecka
model with hyperons dominate. Here one thing should be noted that with
increase in the baryon density, for all the models, the
mean field value of $\rho_{0}^{3}$ increases due to increase in
proton fractions. As a result, the neutron energy and hence the
neutrino energy decreases. After a certain
density, neutrino energy becomes negative and the reaction stops. This
phenomena is more pronounced in case of models without hyperons
at densities greater than $1 fm^{-3}$.
The neutrino emissivity from nonrelativistic model is of the
same order as the relativistic models, whereas, the neutrino
emissivity due to quark URCA processes are in general lower
compared to nucleonic URCA processes.
\par
In fig.4, we have plotted neutrino emissivity
from the reaction $n\rightarrow p~+~e~+\bar \nu_{e}$,
with temperatures, for two different $\Delta p$. The curves for
both exact and approximate results are given. We find that exact
In our calculation we find that our exact neutrino emissivity is
consistently smaller than the approximate results. In fact, when
$\Delta p$ is large compared to the temperature, our numerical result
is almost same as the neutrino emissivity obtained using the analytic
formula (eq. (\ref{eq:redeps})). On the other hand, for small
$\Delta p (\sim T)$, there is a deviation from the analytic
result. The similar results were obtained for quark matter
system\cite{GPS1}. The deviation of approximate from exact
result can be explained in the following ways. In calculating
the approximate formula, the neutrino momentum is neglected in
the delta function. As long as $\Delta p$ is much larger than T,
this approximation gives correct result and neutrino emissivity
varies as $T^{6}$. On the other hand, when $\Delta p$ is small
($\sim T$), an additional power of $T$ may come in the
denominator\cite{GPS1} and neutrino emissivity is nolonger
proportional to $T^{6}$. The other reason for the difference
comes from the factorization of angle and momentum integrals.
The momenta may differ from the corresponding Fermi momenta by T
in the integral. When $\Delta p \sim T$, there are reasons in
momentum space where $Cos{\theta_{pn}}$ ($\theta_{pn}$ is the
angle between proton and neutron ) is greater than $1$, and the
rest of the integrand is not small. Clearly these regions must
be excluded from the integration as these values of
$Cos_{\theta_{pn}}$ are unphysical. If one does not put this
restriction, which happens when one factorizes angle and
momentum integrals, the phase space integral will be
overestimated.
\par
In our earlier section, we have discussed that a strangeness
changing reaction is proportional to $Sin^{2}\theta_{c}$. Hence,
it would have lower neutrino emissivity compared
to the reactions where there is no change in strangeness.
In the present calculation, we find that the neutrino emissivity
from $n$ to $p$ decay and $\Sigma$ to $\Lambda$ decay are larger
($\sim$ order of magnitude)than the neutrino emissivity from $\Lambda$ to
$p$ and $\Sigma$ to $p$ decay, which are strangeness changing
reactions. Overall, the neutrino emissivity from the neutron
decay is higher than all the other decays.
\par
We have studied the neutrino emissivity from hadronic matter for
different parameter sets. It is found that there is only a small
variation in emissivity with incompressibility.
Also, as mentioned earlier, hyperon couplings are varied in our
calculation. With decrease in hyperon couplings, it becomes
energetically favourable to convert nucleons into hyperons as
hyperons do not feel the predominantly repulsive force. As a
result, with decreasing couplings more and more hyperons get
populated. This implies that with decrease in hyperon couplings,
the neutrino emissivity due to the hyperon decay increases.
\par
In our calculation for neutrino emissivity from hadronic matter,
we find that the departure from the usual approximation \cite{GPS1} arises
due to large $T/\Delta p$. The
similar behaviour have been found for quarks also. So it may be
possible to fit the numerically calculated $\epsilon$
($\epsilon_{exact}$) with a function of the form
$\epsilon_{approx.}(f(x))^{-1}$, where $x=T/\Delta p$. The
function $f(x)$ should be such that for small values of $x$ it
should approach unity. In fig.5,
we have plotted $\epsilon_{approx.}/\epsilon_{exact}$ against
$T/\Delta p$ for different decays with different parameters.
We have fitted the above graph
with a function $f(x)~=~ 1~+~ax~+~bx^{2}~+~cx^{3}$,
with $a=-2.5$, $b=100$ and $c=30$ as
obtained in the case of interacting quarks \cite{GPS1}. It is evident
that with an overall multiplicative factor $\sim 1.5$, same
function can describe both hadronic and quark emissivities. This
factor $1.5$ is due to the nonrelativistic approximation, which
is used to calculate approximate formula of neutrino emissivity
in hadronic decays.
\section{{\bf Conclusion}}
In the present paper, we have studied the neutrino emissivity
due to hadronic and quark URCA processes for different models.
We have also considered effect of incompressibility and hyperon
couplings on neutrino emissivity. It is found that relativistic models,
considered here, in general, have nonzero neutrino emissivity and these
are higher compared to neutrino emissivity due to quark URCA
processes. But the
scenario is not so simple in case of nonrelativistic models.
In such models, neutrino emissivity is highly sensitive to the
nature of three-nucleon interactions. In fact in the present
study, we find that only $UV14+UVII$ gives nonzero emissivity.
For relativistic models, neutrino emissivity is more sensitive
to the hyperon couplings than the incompressibilities.
\par
Our calculation shows that as in the case of quark
matter, approximate formula is not valid for hadronic
weak decays when $T/\Delta p$ is large i.e $\Delta p$ is smaller
compared to $T$. An alternative formula can be used to calculate
the neutrino emissivity in such cases for both hadronic and quark matter.
\par
In conclusion, direct Urca processes in
hadronic decays provides an alternate scenario for rapid cooling,
without the necessity of phase transition to quark matter phase
inside neutron stars.
\vfill
\eject

\vfill
\eject
\newpage
\begin{figure}
\caption{Proton fractions for nonlinear Walecka (NW) model,
first three curves are for incompressibility $K=300 MeV$}
\vskip 0.5in
\caption{Proton fractions for (a) DSC model,
(b) CS model, (c) NW model without hyperons, $K=300 MeV$,
(d) NW model with hyperons, $K=300 MeV$ and (e) nonrelativistic
model UV14+UVII}
\vskip 0.5in
\caption{Variation of neutrino emissivity in c.g.s. units
($gm/cm^3/sec.$) with density for $T=0.5 MeV$, (a) DSC model,
(b) CS model, (c) NW model without hyperons, $K=300 MeV$,
(d) NW model with hyperons, $K=350 MeV$, (e) CCD model, (f) bag
model and (g) nonrelativistic model UV14+UVII}
\vskip 0.5in
\caption{Neutron decay emissivity in c.g.s. units ($gm/cm^3/sec.$)
for incompressibility ($K$) $300 MeV$; for
$n_B=0.4 fm^{-3}$
and $\Delta p= 54.53$ (a) Our result  (b) Approximate
result; for $n_B= 1.4 fm^{-3}$ and $\Delta p= 0.50 MeV$ (c)
Our result,  (d) Approximate result}
\vskip 0.5in
\caption{Neutrino emissivity ratio is plotted against $x= T/Delta p
(Deltap= \Delta p)$.
The points corresponds to different hadronic decays. The line
corresponds to the function $f(x)= 1~+~ax~+bx^2~+~cx^3$. $a=-2.5$
,$b=100$ and $c=30$ as obtained in ref.[17] for quarks.}
\end{figure}
\vfill
\eject
\end{document}